\documentclass[12pt]{article}
\usepackage[margin=1in]{geometry}
\usepackage{amsmath,amsfonts}
\usepackage{graphicx}
\usepackage{natbib}
\usepackage{hyperref}
\usepackage[utf8]{inputenc}
\usepackage{authblk}
\usepackage{url}
\usepackage{subcaption}

\title{Modeling Bulimia Nervosa in the Digital Age: The Role of Social Media}

\author[1]{Brenda Murillo}
\author[1]{Fabio Sanchez}

\affil[1]{Escuela de Matem\'atica-CIMPA, Universidad de Costa Rica, Ciudad Universitaria Rodrigo Facio, San Jos\'e, 11501, Costa Rica}

\date{}

\begin{document}

\maketitle

\begin{abstract}
Globalization has fundamentally reshaped societal dynamics, influencing how individuals interact and perceive themselves and others. One significant consequence is the evolving landscape of eating disorders such as bulimia nervosa (BN), which are increasingly driven not just by internal psychological factors but by broader sociocultural and digital contexts. While mathematical modeling has provided valuable insights, traditional frameworks often fall short in capturing the nuanced roles of social contagion, digital media, and adaptive behavior. This review synthesizes two decades of quantitative modeling efforts, including compartmental, stochastic, and delay-based approaches. We spotlight foundational work that conceptualizes BN as a socially transmissible condition and identify critical gaps, especially regarding the intensifying impact of social media. Drawing on behavioral epidemiology and the adaptive behavior framework by Fenichel et al., we advocate for a new generation of models that incorporate feedback mechanisms, content-driven influence functions, and dynamic network effects. This work outlines a roadmap for developing more realistic, data-informed models that can guide effective public health interventions in the digital era.
\end{abstract}

\section{Introduction}

Bulimia nervosa (BN) is a multifaceted disorder characterized by recurring episodes of uncontrollable binge eating, followed by compensatory behaviors such as purging, fasting, or excessive exercise \cite{APA2013}. While its clinical profile is well established, an increasing focus has been placed on the broader social and cultural determinants that contribute to its onset and persistence. Extensive empirical evidence links media-driven ideals of thinness, promoted through both traditional and digital platforms, to disordered eating behaviors, particularly among adolescents and young adults \cite{Stice2002,Levine2006,Perloff2014}. The American Psychiatric Association estimates that up to $10\%$ of adolescent girls exhibit subclinical symptoms of eating disorders, which, although not meeting diagnostic thresholds, still pose significant health risks \cite{suhag2024social}.

\begin{figure}[ht]
    \centering
    \begin{subfigure}[t]{0.48\textwidth}
        \centering
        \includegraphics[width=\linewidth]{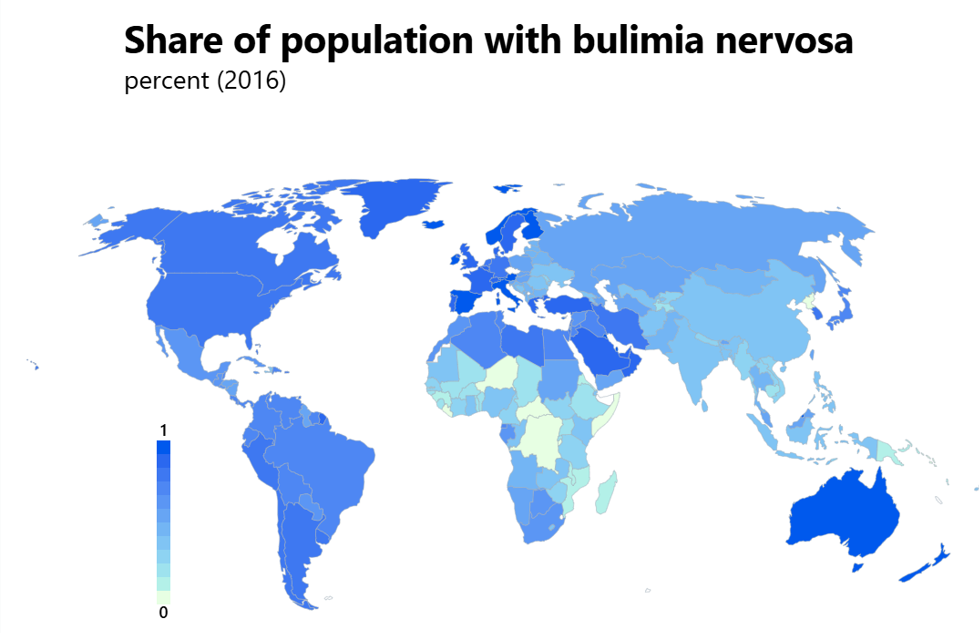}
        \caption{Global distribution of bulimia nervosa prevalence (2016).}
        \label{fig:map}
    \end{subfigure}
    \hfill
    \begin{subfigure}[t]{0.48\textwidth}
        \centering
        \includegraphics[width=\linewidth]{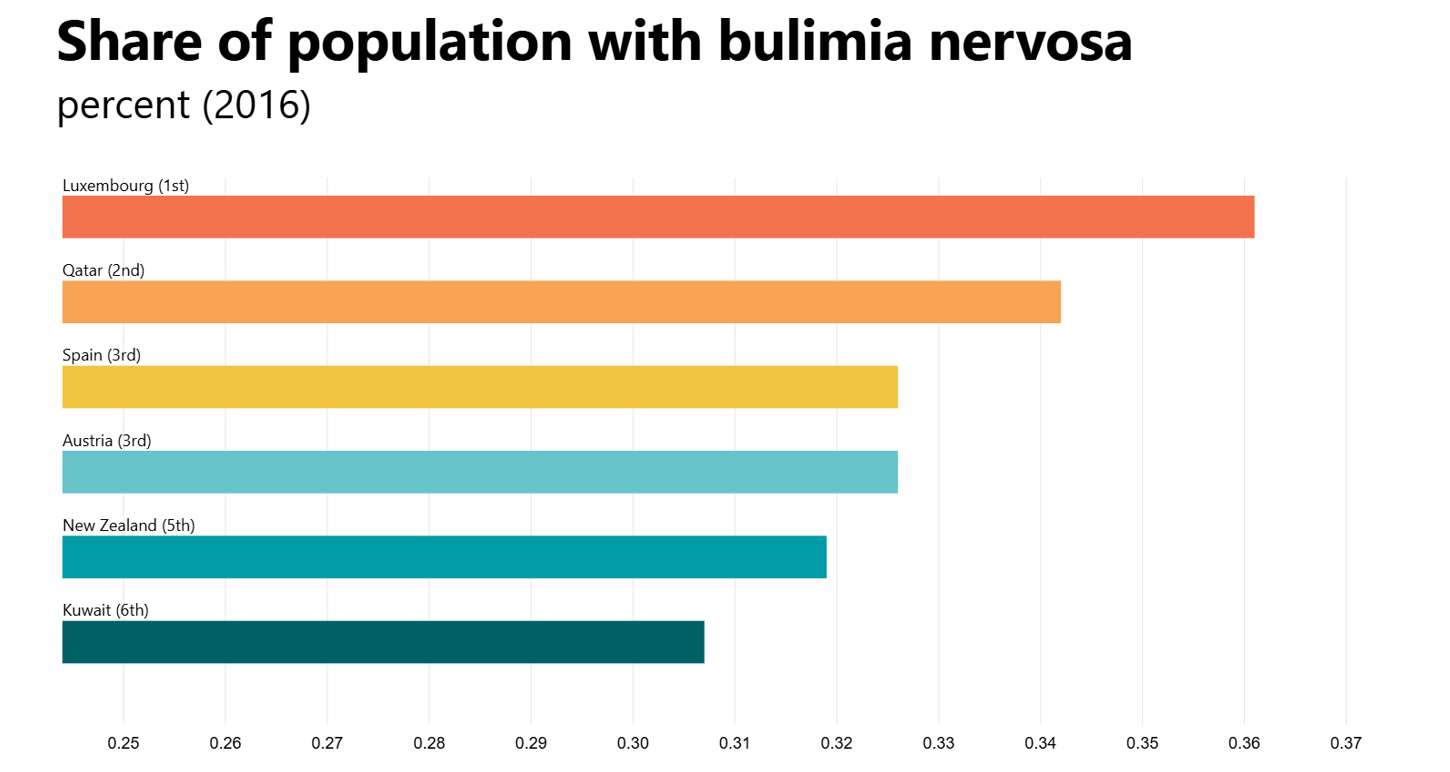}
        \caption{Top five countries by population affected by bulimia nervosa (2016).}
        \label{fig:top_5}
    \end{subfigure}
    \caption{Prevalence of bulimia nervosa across regions and countries. Source: HumanProgress.org.}
    \label{fig:BN_global}
\end{figure}

Understanding the global distribution of BN requires acknowledging regional differences. While eating disorders have historically been more prevalent in Western societies, their incidence is rising in non-Western regions as well. However, research in these contexts remains sparse \cite{suhag2024social,mina}. Figures \ref{fig:map} and \ref{fig:top_5} provide a visual overview of the worldwide prevalence of BN, highlighting the need for context-specific modeling approaches.

From a modeling standpoint, BN can be understood both as a psychological disorder and as a socially transmitted behavior. Parallels with infectious disease models are evident: exposure to triggers (e.g., thin-ideal imagery, peer behavior), varying susceptibility, and progression through behavioral stages align with compartmental modeling structures \cite{gonzalez2003,CastilloChavez2003}. Such models enable the exploration of relapse dynamics, social contagion effects, and the potential impact of targeted interventions, such as awareness campaigns and improved access to treatment.

Gonz\'alez’s seminal model \cite{gonzalez2003} introduced a compartmental approach, dividing individuals into susceptible (S), early-stage bulimic ($B_1$), advanced bulimic ($B_2$), and treated (T) compartments. The transition from S to $B_1$ is influenced by social exposure, analogously to the force of infection in epidemiology. This framework captured both internal reinforcement mechanisms and external cultural pressures.

Modern adaptations must incorporate the influence of social media, which acts as a highly connected network that accelerates behavioral diffusion \cite{Rodgers2016,Fardouly2015}. Platforms such as Instagram and TikTok amplify dissatisfaction with body image and foster echo chambers where disordered eating is normalized through ``pro-ana," ``pro-mia," and ``thinspiration" content. These dynamics create feedback loops in which psychological vulnerability and digital engagement mutually reinforce one another.

As in epidemic modeling, a central challenge is identifying threshold conditions that determine whether BN persists endemically or declines. Tools such as bifurcation analysis and dynamical systems theory are valuable for analyzing these tipping points. Furthermore, introducing heterogeneity in exposure, gender, and socioeconomic context improves model realism and utility.

Approaching BN as a dynamic, socially driven process enables a systems-level understanding that supports theoretically grounded, data-driven intervention strategies.

\section{Modeling the Dynamics of Bulimia Nervosa}

Although BN is widespread and socially mediated, mathematical modeling efforts remain underdeveloped compared to those for classical infectious diseases. Early models translated psychological theories into differential equations to capture behavioral transitions shaped by environmental stimuli.

Gonz\'alez \cite{gonzalez2003} laid the foundation with a compartmental model delineating four classes: susceptible (S), early-stage ($B_1$), advanced ($B_2$), and treated (T). Transitions depended on internal progressions and external social exposure, capturing the essence of behaviorally driven contagion.

Li and Wang \cite{Li2005} extended this by incorporating relapse and recovery dynamics. Their generalized framework introduced threshold parameters akin to the basic reproduction number ($\mathcal{R}_0$), which dictate whether disordered behavior spreads or declines. This formulation was later applied to smoking and alcohol use \cite{Sharomi2011}.

Cho and Kim \cite{Cho2010} proposed a delay differential model, acknowledging psychological response lags to social triggers. This inclusion yielded oscillatory outcomes, paralleling cycles seen in infectious disease spread.

Recent studies emphasize media-driven behavior. Rodgers et al. \cite{Rodgers2016} and Moreno et al. \cite{Moreno2011} documented how online platforms intensify disordered eating through peer validation and algorithmic exposure to curated ideals. While primarily descriptive, their work has set the stage for network-based diffusion modeling.

Ghosh et al. \cite{Ghosh2019} introduced stochastic elements via Markov chains, accounting for randomness and heterogeneity. Such tools pave the way for agent-based simulations capable of evaluating policy interventions under uncertainty.

Analytical methods, such as bifurcation and sensitivity analysis, have also been utilized. González’s work highlighted the possibility of driving the system toward a healthy equilibrium by reducing exposure or enhancing treatment. However, phenomena like multistability or relapse-induced hysteresis remain underexplored.

A critical gap persists in integrating sociocultural data, psychological theory, and dynamical modeling. Future directions must also address comorbidities, gender disparities, and the empirical validation of findings using behavioral and digital data.

\section{Adaptive Behavior and Social Media}

Traditional models often assume static behavior, yet individuals adapt in response to evolving norms and perceived risks. Fenichel et al. \cite{Fenichel2011} formalized this concept in epidemiology, and it is highly applicable to BN.

Here, transition rates from S to $B_1$ depend not only on prevalence but on perceived benefits, risks, and social rewards. Social media serves as a behavioral catalyst, altering perceptions through curated, algorithmically amplified content.

Following Fenichel’s adaptive framework, we define the influence rate:
\[
\lambda(t) = \beta \cdot \frac{B_1(t) + B_2(t)}{P(t)} \cdot \phi(I_t)
\]
where $\beta$ is baseline influence, $P(t)$ is population size, and $\phi(I_t)$ models digital content effects, possibly threshold or saturation-based.

Adaptation may be maladaptive (e.g., increased vulnerability during stress, such as post-holiday dieting trends) or protective (e.g., interventions that decrease the desirability of disordered eating).

Utilities can be defined dynamically:
\[
u_B(t) = U(\text{thin ideal}) - C(\text{health costs}) + S(\text{peer validation})
\]
\[
u_S(t) = U(\text{self-acceptance}) + C(\text{avoiding harm}) + S(\text{social support})
\]
Transitions are guided by which utility dominates, and these evolve in response to media exposure.

Such feedback can cause backward bifurcations or hysteresis \cite{sanchez2005,sanchez2022,calvomonge2023}, meaning recovery may require more substantial efforts than prevention.

Modeling interventions (e.g., media literacy) involves reducing $\phi(I_t)$ or incorporating delays to capture gradual cultural shifts. Agent-based simulations also enable the exploration of peer dynamics and network effects.

This integration transforms static models into adaptive systems that more accurately reflect the complexity of BN in a digital society.

\section{Discussion}

This review consolidates current quantitative approaches to BN, underscoring its nature as a condition that is socially transmitted. Adaptive behavior models, inspired by Fenichel et al. \cite{Fenichel2011}, represent a significant leap toward realism by embedding behavioral feedback into system dynamics.

Social media’s role is no longer peripheral; it is central. Algorithmic amplification shapes exposure and, thus, behavior, necessitating the modeling of $\phi(I_t)$ to assess the effectiveness of interventions.

Utility-based frameworks offer further granularity by modeling individual choices and trajectories under varying pressures. Critically, the presence of hysteresis and backward bifurcation suggests that simply reducing prevalence is insufficient; structural influences must also be addressed.

While foundational models (e.g., \cite{gonzalez2003}) laid essential groundwork, most lack the behavioral flexibility now required. Extensions incorporating network diffusion, delay effects, and heterogeneity are crucial.

Empirical validation remains a significant hurdle. Model accuracy relies on real-time data from various sources, including platforms, surveys, and clinical observations. Interdisciplinary collaboration is key.

The evolution of BN modeling reveals a clear shift toward more integrative and socially aware approaches. Embracing this trajectory will enable the development of tools capable of explaining, predicting, and ultimately intervening in the persistence of BN within contemporary digital culture.


\end{document}